\documentclass[aps,prl,reprint,superscriptaddress,amsmath,amssymb]{revtex4-2}
\usepackage{bm}
\usepackage[pdftex]{graphicx,hyperref}
\hypersetup{colorlinks=true, anchorcolor=blue, linkcolor=blue, citecolor=blue, filecolor=blue, urlcolor=blue}

\begin{document}

\title{Quantum Turbulence Coupled with Externally Driven Normal-Fluid Turbulence in Superfluid $^4$He}

\author{Satoshi Yui}
\email[E-mail:]{syui@keio.jp}
\affiliation{Research and Education Center for Natural Sciences, Keio University, 4-1-1 Hiyoshi, Kohoku-ku, Yokohama 223-8521, Japan}

\author{Hiromichi Kobayashi}
\affiliation{Research and Education Center for Natural Sciences, Keio University, 4-1-1 Hiyoshi, Kohoku-ku, Yokohama 223-8521, Japan}
\affiliation{Department of Physics, Hiyoshi Campus, Keio University, 4-1-1 Hiyoshi, Kohoku-ku, Yokohama 223-8521, Japan}

\author{Makoto Tsubota}
\affiliation{Department of Physics \& Nambu Yoichiro Institute of Theoretical and Experimental Physics (NITEP), Osaka City University, 3-3-138 Sugimoto, Sumiyoshi-ku, Osaka 558-8585, Japan}

\author{Rio Yokota}
\affiliation{Global Scientific Information and Computing Center, Tokyo Institute of Technology, 2-12-1 O-okayama, Meguro-ku, Tokyo 152-8550, Japan}

\date{\today}

\begin{abstract}
The coupled dynamics of quantum turbulence (QT) and normal-fluid turbulence (NFT) have been a central challenge in quantum hydrodynamics, since it is expected to cause the unsolved T2 state of QT.
We numerically studied the coupled dynamics of the two turbulences in thermal counterflow.
NFT is driven by external forces to control its turbulent intensity, and the fast multipole method accelerates the calculation of QT.
We show that NFT enhances QT via mutual friction.
The vortex line density $L$ of the QT satisfies the statistical law $L^{1/2} \approx \gamma V_{ns}$ with the counterflow velocity $V_{ns}$.
The obtained $\gamma$ agrees with the experiment of T2 state, validating the idea that the T2 state is caused by NFT.
We propose a theoretical insight into the relation between the two turbulences.
\end{abstract}

\maketitle

Quantum turbulence (QT) refers to the turbulent state of a superfluid \cite{halperin09,tsubota13,barenghi14}, and it is relevant to a wide range of branches of physics from atomic to cosmological scales, e.g., superfluid $^4$He and $^3$He \cite{vinen06,vinen10}, atomic Bose--Einstein condensates (BECs) \cite{henn09}, neutron stars \cite{packard72}, galactic dark-matter BECs \cite{sikivie09}, and the holographic model \cite{chesler13}.
Although intensive studies on QT have been performed, fundamental questions remain.
The most notable case is the interaction between QT and normal fluid (thermal excitations).
According to the two-fluid model, superfluid $^4$He is understood as an intimate mixture of an inviscid superfluid and a viscous normal fluid \cite{kapitza38,landau41}.
Superfluid and normal fluid move with individual velocities ${\bm v}_s$ and ${\bm v}_n$, respectively.
In a quantum-turbulent state, mutual friction (MF) occurs between the two fluids.
This coupled two-fluid dynamics is an important problem in quantum hydrodynamics.
This Letter directly investigates the fully coupled dynamics of QT and normal-fluid turbulence (NFT).
This coupled dynamics is expected to be the origin of the unsolved state of QT.

This study addresses QT in superfluid $^4$He, which is a typical system of quantum hydrodynamics \cite{donnelly91}.
A quantized vortex has a vortex-filament structure, i.e., its circulation is defined as $\kappa = 1.00 \times 10^{-3} ~{\rm cm^2/s}$ around the thin core of $0.1 ~{\rm nm}$  \cite{feynman55}.
Thermal counterflow refers to an experiment producing QT.
In a closed channel, the normal fluid flows from a heater at the closed end, and the superfluid flows in the opposite direction because of mass conservation.
Using the mean superfluid velocity $V_s=|\langle {\bm v}_s \rangle|$ and normal-fluid velocity $V_n=|\langle {\bm v}_n \rangle|$, the counterflow relation is expressed as $\rho_s V_s = \rho_n V_n$.
Here, $\langle \cdots \rangle$ denotes the spatial average, and $\rho_s$ and $\rho_n$ denote the superfluid and normal-fluid densities, respectively.
Above some critical velocity, QT appears in the form of a tangle of the vortex filaments.
QT is characterized by a vortex line density $L = (1/\Omega) \int_{\mathcal L} d\xi$, which represents the line length of the vortex filaments in a unit volume.
Here, $\xi$ is the arc length of the filaments, and ${\mathcal L}$ denotes the filaments in a sample volume $\Omega$.
By increasing the mean counterflow velocity $V_{ns} = |\langle {\bm v}_n - {\bm v}_s \rangle|$, the value of $L$ increases.
The steady-state relation is shown as
\begin{equation}
  L^{1/2} = \gamma (V_{ns} - V_0),
  \label{eq:vinen_exp}
\end{equation}
where $V_0$ is a practical parameter \cite{vinen57c,tough82,tsubota17}.
The response coefficient $\gamma$ depends on the temperature.

Although extensive studies have been conducted, there remain mysterious phenomena involving QT.
The most notable case is that QT has two different states, T1 and T2, in counterflow, which are characterized by the values of $\gamma$ \cite{tough82}.
The T1 state appears with smaller values of $\gamma_1$ in $V_{c1} < V_{ns} < V_{c2}$, and the T2 state exhibits larger $\gamma_2$ in $V_{c2} < V_{ns}$.
Here, $V_{c1}$ is the critical velocity for the superfluid turbulent transition, and $V_{c2}$ is the critical velocity for the T1--T2 transition.
Some studies have suggested that, in the T2 state, both two fluids are turbulent, whereas the normal fluid is laminar in the T1 state \cite{tough82,melotte98}.
Thus, the T2 state is expected to be caused by coupled dynamics, but its mechanism has not yet been elucidated.

Recent experimental breakthroughs gave us important knowledge of the coupled dynamics.
The experiments visualized the individual dynamics of the two fluids using particle tracking velocimetry \cite{paoletti08,mastracci19,moroshkin19} and particle imaging velocimetry (PIV) \cite{guo09,guo10,marakov15}.
This Letter is interested in the recent PIV experiment of the T2 state \cite{gao17}.
The PIV experiment observed the vortex line density and normal-fluid velocity fluctuations, i.e., the statistical values of both turbulences.
With reference to the PIV experiment, we investigate the T2 state theoretically and numerically.

To elucidate the T2 state, a theoretical study should consider the fully coupled dynamics of the two fluids.
In previous studies, only the superfluid dynamics were investigated using the VFM \cite{schwarz85,schwarz88,adachi10,baggaley15,yui15}, and the Hall--Vinen--Bekarevich--Khalatnikov (HVBK) equations were used for both fluids \cite{bertolaccini17,biferale19,biferale19b,kobayashi19}.
A few pioneering studies on the coupled dynamics of the VFM and HVBK equations have been performed, but for restricted situations \cite{kivotides00,kivotides07}.
In recent years, some studies addressed fully coupled two-fluid dynamics \cite{yui18,yui20,galantucci20}.
By developing the recent method, we analyze the T2 state.

This Letter aims to investigate the fully coupled dynamics of QT and NFT in counterflow to uncover the T2 state.
It is important to determine whether the response coefficient $\gamma$ in the dual turbulent state agrees with the PIV experiment of T2 \cite{gao17}.
The NFT should be caused by large-scale shear stresses from the channel walls.
However, the simulation in this Letter is performed in a periodic cube without any wall effects.
Hence, we obtain a steady state of the NFT using external forces, which correspond to a substitute for the wall shear stresses.
The advantages of this method are as follows: (1) we can control the intensity of the NFT to analyze the dependence of the statistical values and (2) there are no complications near channel walls (e.g., superfluid boundary layer \cite{baggaley15,yui15b}).
We note that the external forces act only on large scales, and the injected energy is transferred from large to small scales.
In addition, we apply a fast multipole method with graphics-processing-unit parallelization to the VFM  \cite{yokota07,yokota09} (see Supplemental Material for details).
This method greatly accelerates the calculation, allowing us to easily obtain sufficient data to investigate the statistical law of the T2 state.

This Letter consists as follows.
First, we introduce the formulations of the coupled dynamics.
We then perform a numerical simulation and confirm that the dual turbulent state becomes statistically steady in the counterflow.
We analyze the vortex line density of QT and the velocity fluctuations of NFT.
In addition, we propose a theoretical insight into the steady-state relationship between these statistical values.
And, we confirm that the vortex line density increases with the velocity fluctuations of the NFT, as expected from the theoretical insight.
Finally, we show that the response coefficient $\gamma$ in the dual turbulent state agrees with the experimental value of T2.

\begin{figure}
  \centering
  \includegraphics[width=1.0\linewidth]{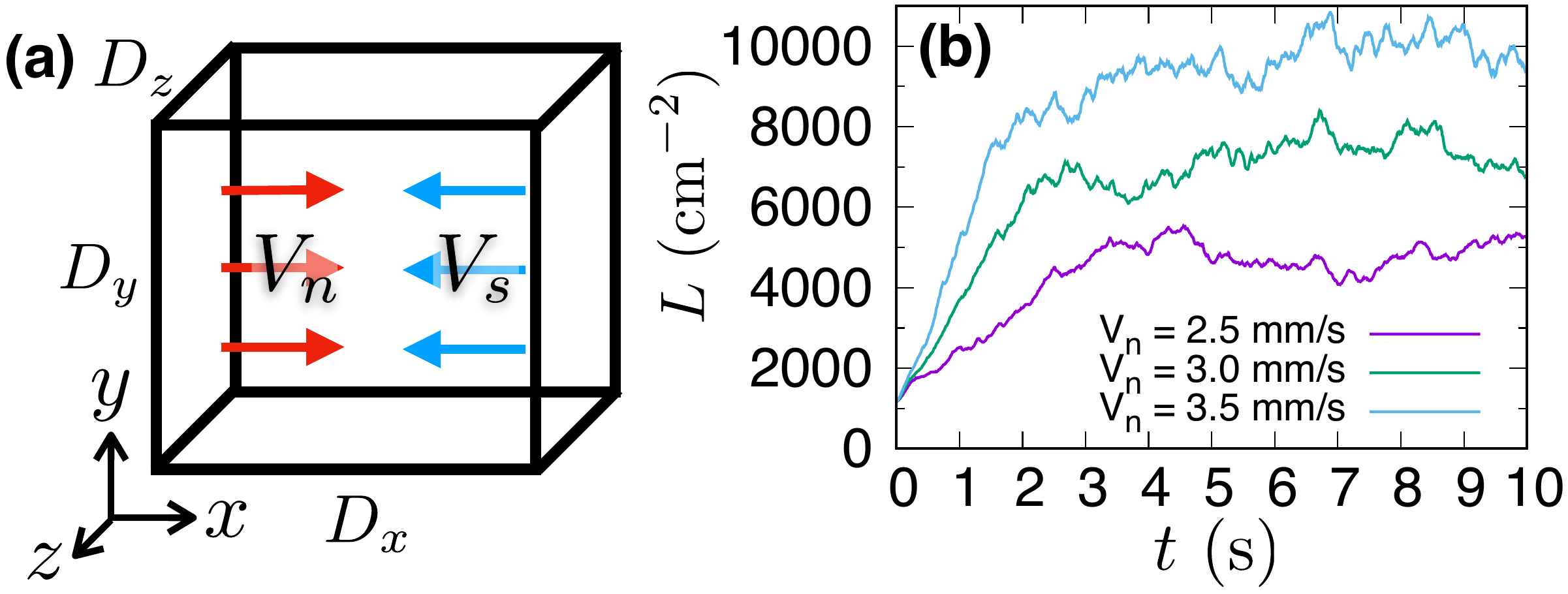}
  \caption{
  (a) Schematic of the numerical simulation of the thermal counterflow.
  (b) Vortex line density as a function of time at $F_0 / \rho_n = 0.30 ~{\rm mm/s^2}$ and $T=1.9 ~{\rm K}$.
  }
  \label{density_time.pdf}
\end{figure}

In the two-fluid model with QT, the two fluids obey the individual dynamics, and they affect each other via MF \cite{yui20}.
The superfluid dynamics is well described by the VFM \cite{schwarz85,adachi10}.
The position vector of the vortex filaments is expressed as ${\bm s}(\xi)$ with filament arc length $\xi$.
The superfluid velocity ${\bm v}_s$ is determined by the Biot--Savart law as ${\bm v}_s ({\bm r}) = (\kappa/4\pi) \int_{\mathcal L} [({\bm s} - {\bm r}) \times d{\bm s}]/|{\bm s} - {\bm r}|^3 + {\bm v}_{s,b} + {\bm v}_{s,a}$.
Here, ${\bm v}_{s,b}$ and ${\bm v}_{s,a}$ denote the velocity induced by the boundary condition and the externally applied velocity, respectively.
The velocity ${d{\bm s}}/{dt}$ of the vortex filaments is given by
\begin{equation}
  \frac{d{\bm s}}{dt} = {\bm v}_s + \alpha {\bm s}' \times {\bm v}_{ns} - \alpha' {\bm s}' \times ({\bm s}' \times {\bm v}_{ns}),
  \label{eq:filament}
\end{equation}
where ${\bm v}_{ns} = {\bm v}_n - {\bm v}_s$ is the local relative velocity of the two fluids \cite{schwarz85}.
The MF terms contain temperature-dependent coefficients $\alpha$ and $\alpha'$.
The normal-fluid velocity ${\bm v}_n$ is governed by the HVBK equations as follows:
\begin{equation}
  \frac{\partial {\bm v}_n}{\partial t} + ({\bm v}_n \cdot \nabla) {\bm v}_n = -\frac{1}{\rho} \nabla p + \nu_n \nabla^2 {\bm v}_n + \frac{ {\bm F}_{ns} +  {\bm F}_{\rm ext} }{\rho_n},
  \label{eq:navier}
\end{equation}
with total density $\rho$, kinetic viscosity $\nu_n$, and effective pressure $p$ \cite{donnelly91}.
The MF with the vortex filaments is contained by ${\bm F}_{ns}({\bm r}) = [1/\Omega'({\bm r})]\int_{{\mathcal L}' ({\bm r})} {\bm f}(\xi) d\xi$, where ${\bm f}(\xi)/\rho_s \kappa = \alpha {\bm s}' \times ({\bm s}' \times {\bm v}_{ns}) + \alpha' {\bm s}' \times {\bm v}_{ns}$ \cite{barenghi83}.
Here, ${\mathcal L}' ({\bm r})$ refers to the filaments in the local subvolume $\Omega' ({\bm r})$ at ${\bm r}$.
In this study, we apply external forces ${\bm F}_{\rm ext}$ to drive the NFT.
The detailed forms are the Arnold--Beltrami--Childress profile \cite{dombre86}: ${\bm F}_{\rm ext} ({\bm r}) = F_0 \sum_{n=1}^2 (B \cos n \tilde{y} + C \sin n \tilde{z}, C \cos n \tilde{z} + A \sin n \tilde{x}, A \cos n \tilde{x} + B \sin n \tilde{y})$, where $\tilde{i} = 2 \pi i /D_i$ for $i = x, y, z$, and $D_i$ denotes the width of the system, as shown in Fig. \ref{density_time.pdf}(a).
The external forces act only on large scales, i.e., the superposition is limited to $n=1,2$.
In this study, the coefficients are prescribed as $A=1$, $B=-1$, and $C=5$ \cite{galantucci20}.
The amplitude $F_0$ is adjusted to vary the intensity of the NFT.
In addition, we use the incompressible condition $\nabla \cdot {\bm v}_n = 0$ for the normal fluid.

The simulations are performed as follows.
The computational volume is $\Omega = D_x D_y D_z = (1.0 ~{\rm mm})^3$, as shown in Fig. \ref{density_time.pdf}(a).
The vortex filaments are discretized into a series of points with the separation $\Delta \xi$, where $\Delta \xi_{\rm min} = 0.008 ~{\rm mm} < \Delta \xi < 0.024 ~{\rm mm}$.
The temporal integration of Eq. (\ref{eq:filament}) is performed using the fourth-order Runge--Kutta method.
When the two vortex filaments approach more closely than $\Delta \xi_{\rm min}$, the filaments are reconnected to each other \cite{adachi10}.
Filaments with lengths of less than $5 \times \Delta \xi_{\rm min}$ are removed \cite{tsubota00}.
The normal fluid is discretized by a homogeneous spatial grid of $N_x N_y N_z = 40^3$, and the spatial resolution is $\Delta x = \Delta y = \Delta z = D_x / N_x$.
The subvolume of the MF is $\Omega' = \Delta x \Delta y \Delta z$.
The temporal integration of Eq. (\ref{eq:navier}) is performed using the second-order Adams--Bashforth method.
The spatial differentiation of Eq. (\ref{eq:navier}) is obtained using the second-order finite-difference method.
The large eddy simulation with the coherent-structure Smagorinsky model is used to contain the turbulent viscosity of the sub-grid scales of the normal fluid \cite{kobayashi05}.
The periodic boundary condition is applied in the $x$, $y$, and $z$ directions.
The initial states are eight vortex filament rings with a radius of $0.23 ~{\rm mm}$ and a laminar normal flow.
The temperature corresponds to $1.9 ~{\rm K}$.

\begin{figure}
  \centering
  \includegraphics[width=1.0\linewidth]{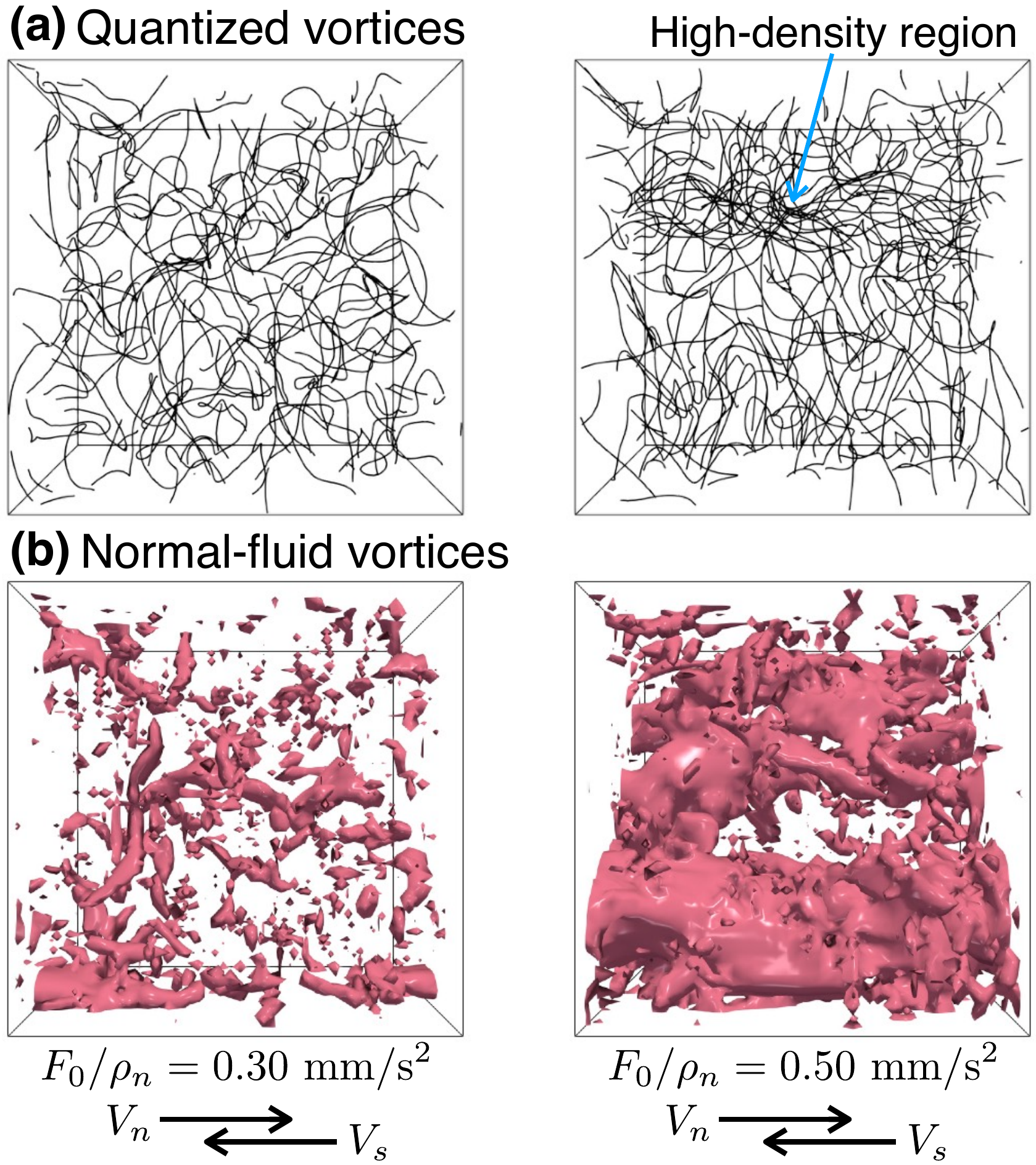}
  \caption{
  QT and NFT in statistically steady states at $V_n = 3.0 ~{\rm mm/s}$.
  (a) Vortex filaments.
  (b) Normal-fluid vortices.
  The surfaces show rotational regions with $Q > 50 ~{\rm s^{-2}}$.
  }
  \label{tangle.pdf}
\end{figure}

QT develops to a statistically steady state in the applied counterflow.
The counterflow is applied, as shown in Fig. \ref{density_time.pdf}(a).
The normal fluid flows in the $x$ direction, where the values of $V_n$ are prescribed in the temporal developments.
Because of the counterflow relation, the applied superfluid velocity is defined as ${\bm v}_{s,a} = - (\rho_n V_n / \rho_s) \hat{\bm x}$, where $\hat{\bm x}$ is the unit vector along the $x$ axis.
Figure \ref{density_time.pdf}(b) shows developments of the vortex line density $L$ at some values of $V_n$, where the NFT occurs with the forcing amplitude $F_0 / \rho_n = 0.30 ~{\rm mm/s^2}$.
The values of $L$ increase from the initial value and fluctuate at some constant values after $t \sim 5 ~{\rm s}$.
This means that QT reached a statistically steady state in the dual turbulent state, where the energy injections and dissipations are statistically balanced.

Before the analysis of the statistical values, we give an overview of the 3D structures of the QT and the NFT in the steady state (see the movie in Supplemental Material).
Figure \ref{tangle.pdf}(a) shows the vortex filaments in the steady states at $V_n = 3.0 ~{\rm mm/s}$.
When the forcing amplitude is $F_0 /\rho_n = 0.30 ~{\rm mm/s^2}$, the vortex tangle tends to be spatially homogeneous, which is similar to the QT obtained in previous studies with prescribed uniform normal-fluid flows \cite{schwarz88,adachi10}.
Figure \ref{tangle.pdf}(b) shows the normal-fluid vortices, i.e., the surfaces show rotational regions with $Q > 50 ~{\rm s^{-2}}$, where $Q$ is the second invariance $Q = (1/2) (\omega_{ij} \omega_{ij} - S_{ij} S_{ij})$ of the velocity gradient tensors with vorticity $\omega_{ij} = (1/2)( \partial v_{n,j} / \partial x_i  - \partial v_{n,i} / \partial x_j )$ and strain $S_{ij} = (1/2) ( \partial v_{n,j} / \partial x_i + \partial v_{n,i} / \partial x_j )$ \cite{hunt88}.
Here, $v_{n,i}$ denotes the $i$ component of ${\bm v}_{n}$.
At $F_0 /\rho_n = 0.50 ~{\rm mm/s^2}$, the external forces intensify the NFT, and larger normal-fluid vortices appear.
The normal-fluid flow is largely inhomogeneous, allowing the inhomogeneous MF to act on the vortex filaments.
Thus, as shown in Fig. \ref{tangle.pdf}(a)(right), the vortex tangle becomes spatially inhomogeneous, with dense filament regions.
These coupled inhomogeneous structures will be characteristic of the T2 state.

\begin{figure}
  \centering
  \includegraphics[width=1.0\linewidth]{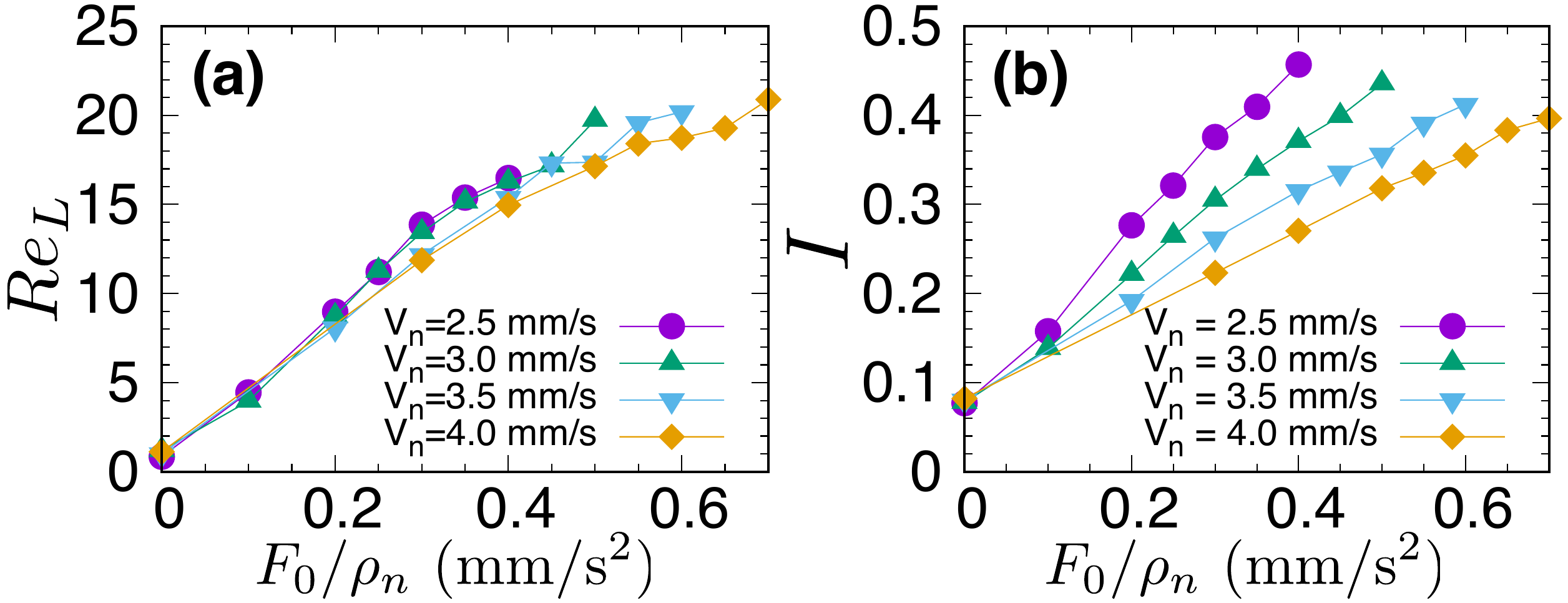}
  \caption{
  (a) Reynolds number with integral length and (b) velocity fluctuations as a function of the amplitude of the external forces.
  }
  \label{reynolds_fluctuation.pdf}
\end{figure}

\begin{figure*}
  \centering
  \includegraphics[width=1.0\linewidth]{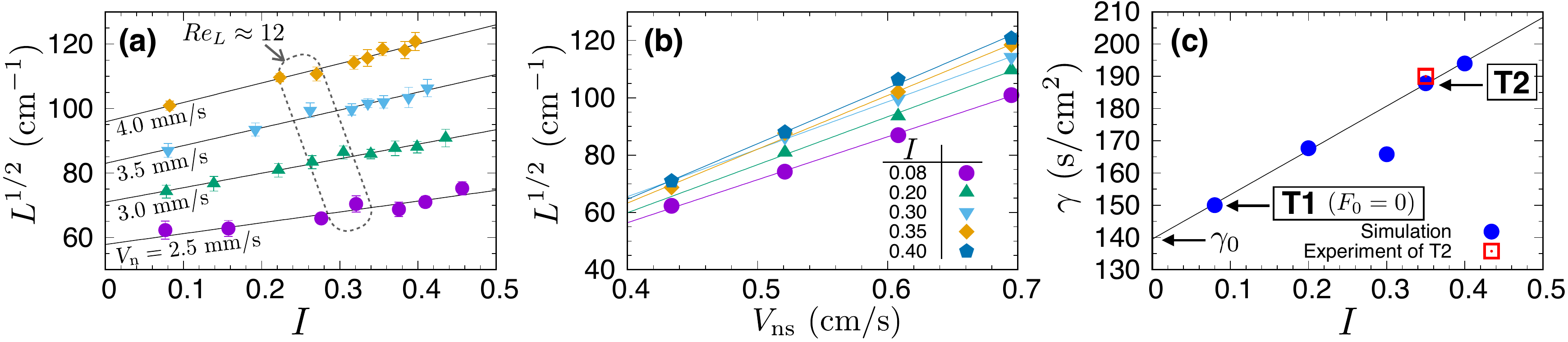}
  \caption{
  Square root of the vortex line density as a function of (a) normal-fluid velocity fluctuations and (b) counterflow velocity.
  (c) Response coefficient $\gamma$.
  The experimental value of the T2 state was obtained from Ref. \cite{gao17}.
  }
  \label{vinen_gamma.pdf}
\end{figure*}

To consider the dual turbulent state, it is important to determine whether the normal fluid is turbulent.
We then use a Reynolds number with the fluctuation velocity $\Delta v_n = \langle ({\bm v}_n - V_n \hat{\bm x})^2 \rangle^{1/2}$ and integral length $J$: $Re_L = \Delta v_n J / \nu_n$.
$Re_L \sim 10$ is known as the critical value of the turbulent transition at large scales.
Figure \ref{reynolds_fluctuation.pdf}(a) shows the values of $Re_L$ at $t=7.5 ~{\rm s}$ in the steady state.
The gradients of the lines tend to change at approximately $Re_L = 12$. 
Thus, the normal fluid should become turbulent at large scales near $Re_L = 12$ as expected.
The PIV experiment \cite{gao17} observed the streamwise velocity fluctuations $I$ of a normal fluid:
\begin{equation}
  I = \frac{\langle (v_{n,x} - V_n)^2 \rangle^{1/2}} {V_n}.
\end{equation}
We analyze the value of $I$ as a statistical value of the NFT.
Figure \ref{reynolds_fluctuation.pdf}(b) shows the values of $I$ averaged over the steady states for $5 ~{\rm s} < t < 10 ~{\rm s}$.
The values of $I$ increase with the forcing amplitude $F_0 / \rho_n$.
The experimental value of the dual turbulent state was $I \approx 0.35$ at $T=1.85 ~{\rm K}$ \cite{gao17}, and we obtained this value in this simulation.
We show that the response coefficient $\gamma$ of the current simulation agrees with the experimental value at $I \approx 0.35$ later.

The statistical value of QT refers to the vortex line density $L$, and that of NFT denotes the normal-fluid velocity fluctuations $I$.
Let us consider the relation between $L$ and $I$.
The Vinen equation describes the development of $L$ in a spatially averaged form \cite{vinen57c}.
To contain the effects of the inhomogeneous structure of the dual turbulent state, we consider the fluctuations $\delta v_{ns} = v_{ns} - V_{ns}$ and $\delta L = L' - L$, where $L'$ denotes the local vortex line density.
Then, the Vinen equation can be expanded to
\begin{equation}
  \frac{dL}{dt} = \chi_1 \alpha \langle v_{ns} (L+ \delta L)^{3/2} \rangle  - \chi_2 \frac{\kappa}{2\pi} L^2,
  \label{eq:vinen_ex}
\end{equation}
where $\chi_1$ and $\chi_2$ are the temperature-dependent coefficients.
The correlation term $\left \langle v_{ns} (L+\delta L)^{3/2} \right \rangle$ can be rewritten as
\begin{eqnarray}
  && \left \langle (V_{ns} + \delta v_{ns}) L^{3/2} \left( 1 + \frac{3}{2} \frac{\delta L}{L} + \cdots \right) \right \rangle \nonumber \\
  &\approx& V_{ns} L^{3/2} + \frac{3}{2} L^{1/2} \left \langle \delta v_{ns} \delta L \right \rangle.
\end{eqnarray}
Here, the terms with $\langle \delta L \rangle$ and $\langle \delta v_{ns} \rangle$ vanish because $\langle \delta L \rangle = \langle \delta v_{ns} \rangle = 0$.
In addition, the higher-order terms with $\langle \delta v_{ns} \delta L^2 \rangle$, $\langle \delta L^3 \rangle$, $\cdots$ are ignored because they are small.
Thus, we obtain the steady-state relation from Eq. (\ref{eq:vinen_ex}):
\begin{equation}
  L^{	1/2} = \gamma_0 \left( V_{ns} + \frac{3}{2} \left \langle \delta v_{ns} \frac{\delta L}{L} \right \rangle \right),
\end{equation}
where $\gamma_0 =  \chi_1 2 \pi \alpha / (\chi_2 \kappa)$.
By writing the correlation as $C_{vL} = \langle \delta v_{ns} \delta L \rangle / ( \langle \delta v_{ns}^2 \rangle ^{1/2} \langle \delta L^2 \rangle ^{1/2} )$ and the fluctuations of $L'$ as $ I_q = \langle \delta L^2 \rangle^{1/2} / L$, the correlation term can be expressed as $\left \langle \delta v_{ns} \delta L / L \right \rangle = C_{vL} \langle \delta v_{ns}^2 \rangle^{1/2} I_q$.
We put $\langle \delta v_{ns}^2 \rangle^{1/2} = k \langle \delta v_n^2 \rangle^{1/2}$ with the parameter $k$ and obtain $ \langle \delta v_{ns}^2 \rangle^{1/2} = k V_n I = k (\rho_s/\rho) V_{ns} I$.
Therefore, the expanded relation of the steady state is obtained as
\begin{eqnarray}
  L^{1/2} &=&  \gamma (V_{ns} - V_0), \label{eq:vinen_expand}\\
  \gamma &=& \gamma_0 \left( 1 + \epsilon I \right), \label{eq:vinen_expand_gamma}
\end{eqnarray}
where $\epsilon = (3/2) C_{vL} k (\rho_s / \rho) I_q$.
Here, the practical parameter $V_0$ is added.
The steady-state relationship does not change from Eq. (\ref{eq:vinen_exp}), but the coefficient $\gamma$ increases with $I$.
When the correlation $\epsilon I$ is small, $\gamma$ is close to $\gamma_0$.
We note that $\gamma_0 \ne \gamma_1$ because in the T1 state, the values of $I$ are small but not negligible \cite{mastracci19,yui20}.

We now analyze the statistical values obtained by the simulation.
Figure \ref{vinen_gamma.pdf}(a) shows the values of $L^{1/2}$ averaged over the steady states for $5 ~{\rm s} < t < 10 ~{\rm s}$.
The obtained values of $L^{1/2}$ increase with $I$, because the NFT significantly enhances the QT when $I$ is larger.
From Eqs. (\ref{eq:vinen_expand}) and (\ref{eq:vinen_expand_gamma}), $L^{1/2}$ will be proportional to the normal-fluid velocity fluctuations $I$.
The results tend to obey $L^{1/2} \propto I$ as expected.
The values tend to deviate from the fitted solid lines near $Re_L = 12$.
This may be due to the turbulent transition of the normal fluid occurring near $Re_L = 12$.

We confirm that the steady-state relation in Eq. (\ref{eq:vinen_expand}) is satisfied even in the dual turbulent state, which has never been confirmed numerically.
The PIV experiment showed that the velocity fluctuations $I$ did not depend on $V_{ns}$ \cite{gao17}.
Thus, we investigate the relationship by selecting the data from Fig. \ref{vinen_gamma.pdf}(a), which are close to some constant values of $I$.
Figure \ref{vinen_gamma.pdf}(b) shows the mean values of $L^{1/2}$ as a function of $V_{ns}$.
The results satisfy the steady-state relation in Eq. (\ref{eq:vinen_expand}) for different values of $\gamma$.

Finally, we compare $\gamma$ with the experimental value of the T2 state.
Figure \ref{vinen_gamma.pdf}(c) shows the response coefficient $\gamma$ as a function of $I$.
The values are obtained from the slopes of the fitting lines in Fig. \ref{vinen_gamma.pdf}(b).
The experimental value of $I$ is approximately $0.35$ in the T2 state \cite{gao17}.
At $I = 0.35$, the obtained $\gamma$ agrees with the experimental value $\gamma_2 \approx 190 ~{\rm s/cm^2}$ of the T2 state \cite{gao17} (because there is no experimental value at $1.90 ~{\rm K}$ in Ref. \cite{gao17}, we interpolated that visually).
This agreement supports the idea that the T2 state corresponds to the dual turbulent state of the two fluids.
At $I = 0.08$, the obtained value of $\gamma = 150 ~{\rm s/cm^2}$ corresponds to $\gamma_1$ of the T1 state because the forcing amplitude $F_0$ is zero.
In addition, the obtained results tend to satisfy the relation of Eq. (\ref{eq:vinen_expand_gamma}).
The solid line shows the fitted line without the deviated value at $I = 0.30$, and the parameters are $\gamma_0 = 140 ~{\rm s/cm^2}$ and $\epsilon = 0.98$.
The value of $\gamma_0$ agrees with $\gamma = 140 ~{\rm s/cm^2}$ of the simulation \cite{adachi10} with the prescribed uniform flow of normal fluid.
The deviation of $\gamma$ at $I=0.30$ may be caused by the turbulent transition that should occur near $I = 0.30$.

In summary, this study investigated the dual turbulent state of a two-fluid model in a superfluid $^4$He.
This state has been expected as the unsolved T2 state of QT.
Using the developed numerical simulation, we analyzed the statistical values of QT and NFT in the counterflow.
We showed that the vortex line density increased with normal-fluid velocity fluctuations.
By expanding the Vinen equation, we then proposed a steady-state relation between the statistical values of the two fluids.
Our results for the dual turbulent state agreed with the experiment of the T2 state \cite{gao17}; therefore, we should succeed in numerically obtaining the T2 state.
Henceforth, detailed features will be studied, e.g., the energy spectra of the dual turbulent state and temperature dependence.
The agreement with the experiments validates our simulation and theoretical insight, allowing us to pave the way to the frontier of quantum hydrodynamics of the coupled two-fluid model.
For instance, the current method will be applied to uncover the mechanism of the T1-T2 transition.
Similar coupled dynamics could occur in other fields of quantum hydrodynamics and multi-component systems, and some universality may be investigated over a wide range.

\begin{acknowledgments}
S. Y. acknowledges support from a Grant-in-Aid for JSPS Fellow (Grant No. JP19J00967).
H. K. acknowledges the support from JSPS KAKENHI (Grant No. JP18K03935).
M. T. acknowledges the support from JSPS KAKENHI (Grant No. JP20H01855).
\end{acknowledgments}

\bibliography{biblio.bib}
\bibliographystyle{apsrev4-2}

\end{document}


\title{Supplemental Material}

\begin{abstract}
This Supplemental Material details the application of the fast multipole method to the vortex filament model.
\end{abstract}

\maketitle

\section{Application of a fast multipole method to the vortex filament model}
\begin{figure}[b]
  \centering
  \includegraphics[width=0.33\linewidth]{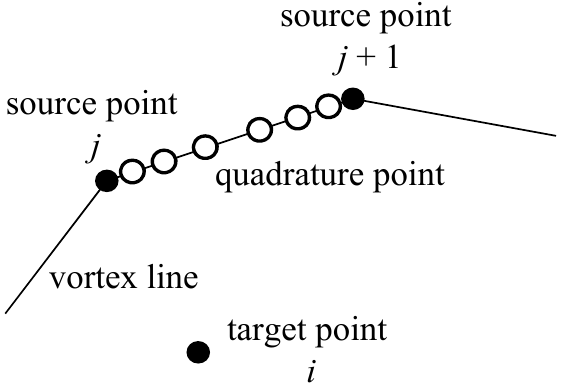}
  \caption{
  Schematics of discretization of the line segment (solid circle: vortex point, open circle: quadrature point).
  }
  \label{fmm.pdf}
\end{figure}

In the numerical simulation of the vortex filament model (VFM) \cite{schwarz85,adachi10}, a vortex filament is discretized into finite straight-line segments.
The line segment has points $j$ and $j+1$ at both ends, as shown in Fig. \ref{fmm.pdf} (we call them vortex points).
In the conventional VFM, the superfluid velocity at a target point $i$ induced by the line segment is obtained from the analytical integration of the Biot--Savart (BS) law along the line segment between $j$ and $j+1$.
The full BS integral \cite{adachi10} is obtained by the summation of the analytical integrations for all line segments.

The basic idea of the fast multipole method (FMM) for the VFM is as follows.
In our application of the FMM, the BS integral along the line segment between $j$ and $j+1$ is performed using the quadrature method.
We place quadrature points that obey the Gauss--Legendre formula on a filament to imitate the source filament, as shown in Fig. \ref{fmm.pdf}.
The superfluid velocity at the target point ${\bm r}_i$ is then approximated by
\begin{equation}
  {\bm v}_{s}({\bm r}_i) \approx \kappa \sum_{q=1}^{M} ({\bm s}'_q \times \nabla G_{iq}) w_q.
  \label{eqs:velocity_2}
\end{equation}
where the summation is taken over all quadrature points, $M$ is the number of all quadrature points, $\kappa$ is the quantized circulation of superfluid velocity, $w_q$ is the Gauss--Legendre weight including the length of the line segment, and ${\bm s}'_q$ is the unit tangent vector along the vortex filaments at the source point ${\bm s}_q$.
Here,
\begin{eqnarray}
  G_{iq} = \frac{1}{4\pi R_{iq}}
\end{eqnarray}
is Green's function, where ${\bm R}_{iq} = {\bm r}_i - {\bm s}_q$.
In the FMM \cite{yokota07,yokota09}, Green's function is approximated using the multipole expansion
\begin{equation}
  \sum_{q=1}^{M} G_{iq} \approx \frac{1}{4\pi} \sum_{n=0}^p \sum_{m=-n}^n r_i^{-n-1} Y_n^m(\theta_i, \phi_i) \sum_{q=1}^{M} \rho_q^n Y_n^{-m} (\alpha_q,\beta_q).
\end{equation}
and the local expansion
\begin{equation}
  \sum_{q=1}^{M} G_{iq} \approx \frac{1}{4\pi} \sum_{n=0}^p \sum_{m=-n}^n r_i^{n} Y_n^m(\theta_i, \phi_i) \sum_{q=1}^{M} \rho_q^{-n-1} Y_n^{-m} (\alpha_q,\beta_q),
\end{equation}
where $p$ is the maximum order of the expansions and the accuracy of the FMM is tunable with $p$.
Here, $(r_i,\theta_i,\phi_i)$ and $(\rho_q,\alpha_q,\beta_q)$ represent the positions ${\bm r}_i$ and ${\bm s}_q$ in spherical coordinates, respectively,
and $Y_n^m$ denotes the spherical harmonics.

The FMM divides the computational box into cells with tree structures \cite{yokota07,yokota09}.
The FMM directly deals with the BS law between a target point and a source point in the same cell, whereas in far fields, the FMM calculates the interaction between cells.
The FMM in this study calculates the direct BS law for the neighboring 1000 vortex points.
This condition is suitable for parallelization in a typical graphics processing unit (GPU). (The suitable condition for the CPU is 50 vortex points per cell.)
To maintain the same level of accuracy as the conventional VFM, we used $p=10$ for the expansion of the FMM to consider the influence of the source filament on a target point.
Eight quadrature points for each line segment were sufficient to obtain the induced velocity within an error of $1 \times 10^{-5}$ between the conventional VFM and the VFM with the FMM.

\begin{figure}[t]
  \centering
  \includegraphics[width=0.42\linewidth]{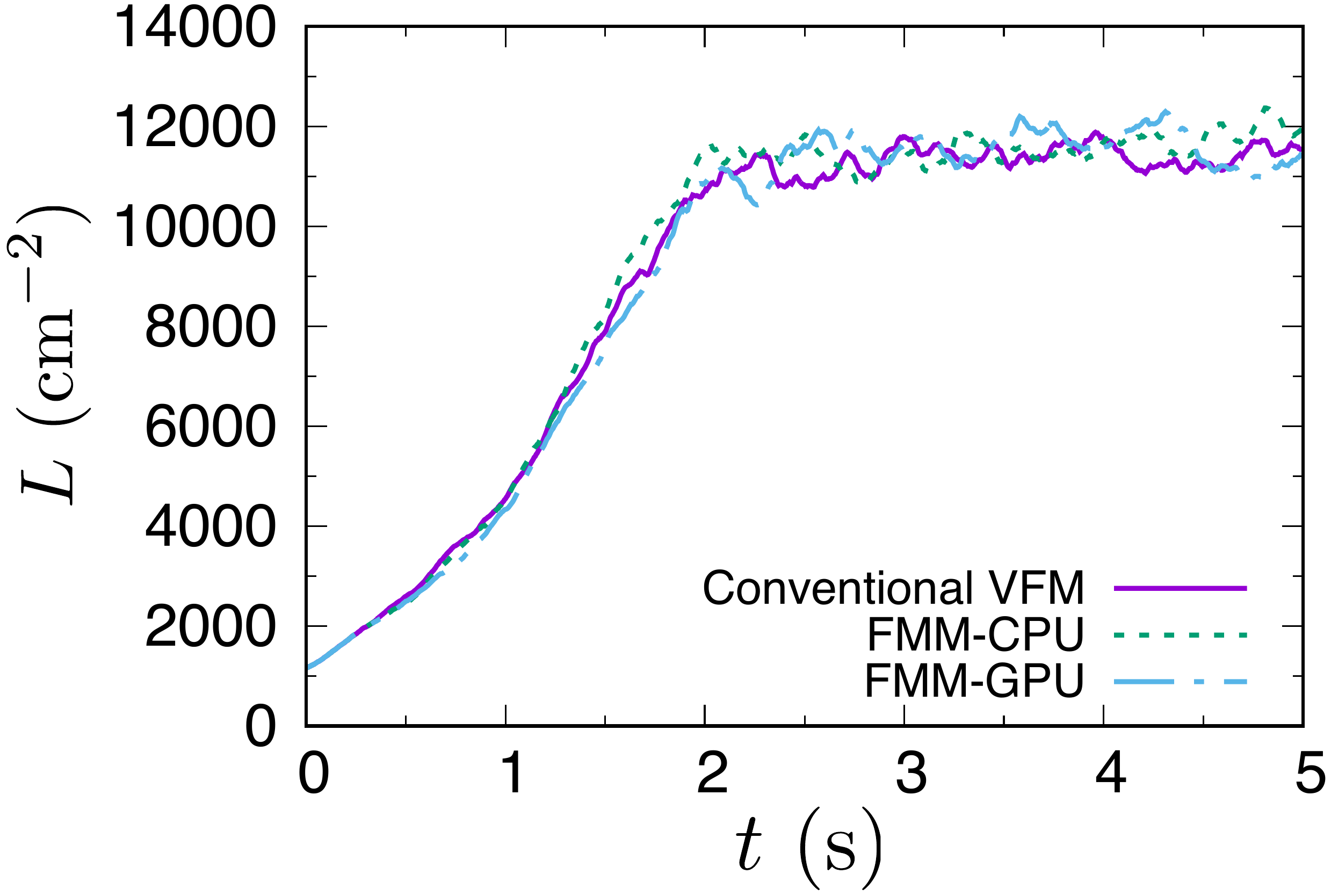}
  \caption{
  Vortex line density in the thermal counterflow as a function of time.
  Conventional VFM: VFM calculation with full BS integration; FMM-CPU: FMM calculation with a single CPU; FMM-GPU: FMM calculation with a single GPU.
  }
  \label{density_fmm.pdf}
\end{figure}

The integration of the BS law diverges at $i = j$ because of the singularity.
In the conventional VFM, the BS law is integrated to avoid singularity for $i = j$.
The velocity induced by the line segment at $i = j$ is also calculated using the local induction approximation (LIA) \cite{schwarz88}.
In the VFM with FMM, the BS integration is performed between a target point $i$ and the quadrature points of the source points.
The vortex point $j$ is not used as a source point, so that a singular point never appears.
In addition, double counting between the LIA term and the BS integration does not occur because the BS integration for the neighboring filaments around the target point at $i = j$ is automatically calculated to be zero
\footnote{
In the same segment, the vector between the target point $i$ and a quadrature point parallels that between the target point $i$ and another quadrature point; thus, the cross-product in the BS law becomes zero.}.
We use $3^3 \times 3^3 \times 3^3 - 1$ periodic boxes surrounding the main computational box to reflect the periodic boundary condition.
If a line segment crosses a periodic boundary, the quadrature points outside the main box are moved by the width of the system size and returned to the main box.

We confirmed that the FMM calculation has sufficient accuracy in a typical quantum-turbulence simulation.
We performed simulations of a thermal counterflow using (1) the conventional VFM, (2) the FMM with a single CPU, and (3) the FMM with a single GPU.
Figure \ref{density_fmm.pdf} shows the obtained values of the vortex line density $L$ as a function of time.
The mean normal-fluid velocity is $V_n = 8.0 ~{\rm mm/s}$, the temperature is $T=1.6 ~{\rm K}$, and the resolutions of the VFM are the same as those of the main document of this Letter.
The $3 \times 3 \times 3 - 1$ periodic boxes were used for the conventional VFM.
The obtained values of the FMM calculations with a single CPU and a single GPU are close to those of the conventional VFM.

The FMM largely accelerates the calculation as follows.
The computational cost of the FMM is known as ${\mathcal O}(N)$ for $N$ points, in contrast to ${\mathcal O}(N^2)$ of the conventional VFM.
In the present study, for $N$ target points and $8N$ source points, the computational cost of the FMM is ${\mathcal O}(\sqrt{N \times 8N}) = {\mathcal O}(2\sqrt{2}N)$.
In the simulations shown in Fig. \ref{density_fmm.pdf}, we compare the calculation time of the FMM with the conventional VFM.
The number of vortex points reached approximately 10,000 in the statistically steady state.
The conventional VFM was calculated over 32 days.
Meanwhile, the calculation of the FMM with a single GPU required only 8 h.
An approximately 100 times faster computation was achieved in this case.
The FMM code applied to this calculation realized a computation using 4096 GPUs with 74\% parallel efficiency for a simulation of classical homogeneous isotropic turbulence using a vortex method with 69 billion particles \cite{yokota13}.

\bibliography{biblio_sup.bib}
\bibliographystyle{apsrev4-2}